\documentclass[useAMS,usenatbib,usenatNishiNishibib,usegraphicx]{mn2e}
\usepackage{times}

\title[The pair of galaxies NGC 6907 and 6908]{H~{\sc i} aperture synthesis 
and optical observations of the pair of galaxies NGC 6907 and 6908}

\author[Scarano Jr. et al.]{S. Scarano Jr. $^1$\thanks{E-mail: scarano@astro.iag.usp.br},
 Felipe R. H. Madsen $^2$, Nirupam Roy $^3$, J. R. D. L\'{e}pine $^1$\\
$^1$ Instituto de Astronomia, Geof\'{\i}sica e Ci\^encias Atmosf\'ericas da
Universidade de S\~ao Paulo, Cidade Universit\'aria\\ 05508-900 S\~ao Paulo, SP, Brazil \\
$^2$ Instituto Nacional de Pesquisas Espaciais, Jd. Granja 12227-010, S\~ao Jos\'e dos
Campos, SP, Brazil \\
$^3$NCRA-TIFR, Post Bag 3, Ganeshkhind, Pune 411 007, India \\}

\begin{document}
\maketitle

\begin{abstract}

NGC 6908, a S0 galaxy situated in direction of NGC 6907, was only recently recognized as a
distinct galaxy, instead of only a part of NGC 6907. We present 21 cm radio synthesis
observations obtained with the GMRT and optical images and spectroscopy obtained with the
Gemini North telescope of this pair of interacting galaxies. From the radio observations we
obtained the velocity field and the H~{\sc i} column density map of the whole region
containing the NGC 6907/8 pair, and by means of the Gemini multi-object spectroscopy we
obtained high quality photometric images and $5 {\AA}$ resolution spectra sampling the two
galaxies. By comparing the rotation curve of NGC 6907 obtained from the two opposite sides
around the main kinematic axis, we were able to distinguish the normal rotational velocity
field from the velocity components produced by the interaction between the two galaxies.
Taking into account the rotational velocity of NGC 6907 and the velocity derived from the
absorption lines for NGC 6908, we verified that the relative velocity between these systems
is lower than 60 km s$^{-1}$. The emission lines observed in the direction of NGC 6908, not
typical of S0 galaxies, have the same velocity expected for the NGC 6907 rotation curve. Some 
emission lines are inside their equivalent absorption lines, reinforcing the idea they were 
not formed in NGC 6908. Finally, the H~{\sc i} profile exhibits details of the interaction,
showing three components: one for NGC 6908, another for the excited gas in the NGC 6907 disk
and a last one for the gas with higher relative velocities left behind NGC 6908 by dynamical
friction, used to estimate the time when the interaction started in $(3.4 \pm
0.6)\times10^7$ years ago.

\end{abstract}


\section{Introduction}

\par Searching for companions among the nearby barred galaxies from the Shapley-Ames Catalog,
Garc\'ia-Barreto et al. (2003) identified NGC 6908 as a discoidal companion of NGC 6907, 10
kpc away from it. They estimated this distance based only on their angular separation. But
according to these authors, the alignment of the objects could be a mere effect of
geometrical superposition, since the narrow band H$\alpha$ images of NGC 6907 did not reveal any
emission in direction of NGC 6908 (Garc\'ia-Barreto et al. 1996). Surace et al. (2004)
recognized NGC 6907 and 6908 as a potential interacting system,  but the low resolution of
their far-infrared images did not allow any further conclusion.

\par In a recent paper, Madore et al. (2007) revealed that NGC 6908, once classified as a PofG
(part of a galaxy), is in fact a lenticular S0(6/7) galaxy, as indicated by its exponential
brightness profile observed  in several filters. Based on long-slit observations, these
authors found different radial velocities for NGC 6908 lines in absorption (3113 $\pm$ 73 km
s$^{-1}$) and in emission (3060 $\pm$ 16) km.s$^{-1}$. Comparing with the radial velocity
that they estimated for NGC 6907 (3190 $\pm$ 5 km s$^{-1}$), the velocity difference between
the two systems would be in the interval of 77 km s$^{-1}$ to 130 km s$^{-1}$. Summing the
evidences of morphological asymmetry  in the arm structure of NGC 6907, the behavior of its
dust lanes and the tail of debris observed at low surface brightness, they conclude that the
two galaxies are certainly interacting.

\par Despite the convincing arguments in favor of the interacting nature of the two galaxies,
one could still question whether the absorption and emission lines observed in the direction
of NGC 6908 are indeed from that galaxy or originate from the rotating disk of NGC 6907.
Studies of rotation curves of spiral galaxies based on ionized gas (emission lines) and on
the stellar component (absorption lines) have shown that, for most galaxies, the rotation of
the two components is similar, but local differences of velocities up to 100 km s$^{-1}$ can
occur (Saito et al. 1984; Fillmore et al. 1986; Kormendy \& Westpfahl 1989; Vega
Beltr\'an et al. 2001; Pizzella et al. 2004 and Rhee et al. 2004). On the other hand, only
luminous S0 galaxies are expected to present detectable emission lines. Since NGC 6908 is a
low luminosity S0(6/7) galaxy (M$_{B}$=-17.4 , Madore et al. 2007) superimposed on the
spiral arm of NGC 6907, the lines used to measure its velocity could well originate in the
rotating material of the disk of NGC 6907.

\par We present in this paper new observations, at radio and optical wavelengths, that
allow us to definitely settle this question, and to propose a scenario that describes in
more detail the recent collision between the two galaxies. We performed high resolution
radio observations  at 21 cm with the Giant Meterwave Radio Telescope (GMRT), and a detailed
spectroscopic study over the field of the system NGC 6907 and 6908, with the Gemini North
telescope.

\par At 21 cm, using the aperture synthesis technique, the H~{\sc i} intensity maps and the
velocity fields of both objects were obtained. The emission of the two objects could be
distinguished and their velocities precisely measured. Furthermore the excess in velocity in
the velocity field was used to estimate the time passed from the start of the interaction.

\par The optical multi-object spectroscopy, employing the GMOS instrument, allowed the
study of the emission and absorption lines of selected H~{\sc ii} regions in the NGC 6907
disk and the nucleus of both galaxies. Each observed spectrum could be compared with its
radio counterpart and the differences in absorption and emission velocities in direction of
NGC 6908 could be explained by the idea that the emission lines were generated on the NGC 6907
disk and the absorption lines were formed on NGC 6908.

\section{The Field of NGC 6907/8}

\begin{figure}
\includegraphics[width=86mm]{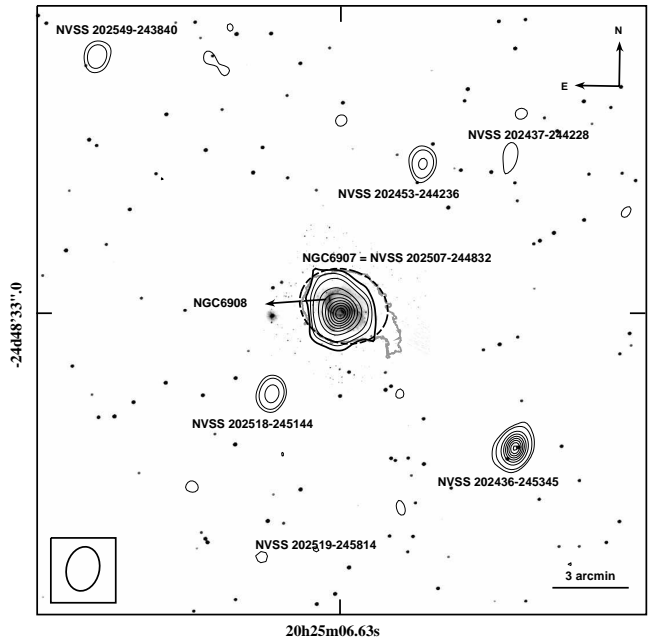}
\caption{Optical image of the field centered in NGC 6907 galaxy and covering $24\times24$
arcsec$^2$. It is overlaid by the GMRT continuum observation. The contours are associated
with the levels 2, 4, 11, 18, 25, 33, 40, 47, 54 and 61 mJy/beam. The external thicker
contour corresponds to the limit of detection of 2 sigmas of the noise level for the
combined continuum image ($\sim$  0.6 mJy/beam). Optical isophotal contour observed at the
Gemini Telescope (gray) and elliptical fit (dashed line) are presented too.} \label{fig1}
\end{figure}

\par Inside a field of $24^{\prime} \times 24^{\prime}$ centered on NGC 6907 (Figure \ref{fig1}), only two extragalactic objects dominate the optical emission:
NGC 6907 itself, an SBbc spiral galaxy (Eskridge et al., 2002) with an absolute B magnitude
M$_{B}$ = -21.34 and a surface brightness of 22.29 mag arcsec$^{-2}$ at half B total light
(Lauberts \& Valentijn 1989) and NGC 6908, a S0(6/7) galaxy with M$_{B}$=-17.4 (Madore et al.
2007), 43.6 arcsec north-west from NGC 6907. Garcia (1993) mentioned NGC 6907 as the main
member of a sparse group of 3 galaxies (IC 4999 and IC 5005, without considering NGC 6908),
also identified in the catalog of galaxy groups of Maia et al (1989), but these objects
cannot be seen in the field shown in the figure. In the radio continuum at 1.4 GHz, NGC 6907
is also the brightest object, as can be seen from the NRAO VLA Sky Survey (Condon et al.
1998). The main properties of NGC 6907/8 galaxies are listed in Table \ref{tbl-1}.

\begin{table}
 \caption{Main properties of the galaxies NGC 6907 and 6908. The index $opt$ is for
the optical observations (combining the g$^{\prime}$ and r$^{\prime}$ filters) and $rad$ for
the radio observations. $RA$ and $Dec$ are the equatorial coordinates, $V_{0}$ is the
systemic velocity, $a$ is the size, $i$ the inclination, $PA$ the position angle, $\mu_{B}$
the surface brightness, $Type$ the morphological type, $D$ the galactocentric Hubble flow
distance, $S_{HI}$ the 21-cm Flux density, $A_{B}$ the Galactic extinction in B-band,
$V_{max}$ the maximum velocity rotation, $r^{\prime}$-$g^{\prime}$ the color in these
filters, $B$ the absolute B-band magnitude, $L_{B}$ the total B luminosity, $M_{dyn}$ the
dynamical mass and $M_{HI}$/L$_{B}$ is the $H~{\sc i}$ mass to B luminosity ratio. The
number for the references are: (1) Eskridge et al. (2002), (2) Madore et al. (2007) and (3)
Amores \& L\'epine (2005).}
 \label{tbl-1}
\begin{tabular}{lll}
\hline
{Parameters} & {NGC 6907} & {NGC 6908} \\
\hline
RA (J2000)& 20$^{h}$25$^{m}$06$^{s}$.63 & 20$^{h}$25$^{m}$08$^{s}$.97 \\ %
Dec (J2000) & -24$^{\circ}$48$^{\prime}$33$^{\prime\prime}$.0 & -24$^{\circ}$48$^{\prime}$04$^{\prime\prime}$.3    \\
$V_{0}$ [km/s] & 3182.4$ \pm $3.9 & 3147$ \pm $21  \\
$a_{opt}$ [arcsec] & 90.0$ \pm $1.2 & 10.3$ \pm $0.6 \\
$a_{rad}$ [arcsec] & 211.8$ \pm $1.9 & -  \\
$i_{opt}$ [$\degr$] & 31.1$ \pm $3.0 & 74.7$ \pm $3.1  \\ %
$i_{rad}$ [$\degr$] & 46.4$ \pm $3.1 & -  \\
$PA_{opt}$ [$\degr$] & 76.4$ \pm $4.1 & 2.8$ \pm $1.4   \\
$PA_{rad}$ [$\degr$] & 52.3$ \pm $11.5 & -   \\
$\mu_{B}$ & 22.36$ \pm $0.05 & 20.81$ \pm $0.08  \\
$Type$ & SBbc $^{(1)}$ & S0(6/7)$^{(2)}$  \\ %
$D$ [Mpc] & 44.5$ \pm $3.1 & 43$ \pm $16  \\
$S_{HI}$ [mJy.km/s] & 48.3$ \pm $1.8 & -   \\
$A_{B}$ & 0.12$ \pm $0.07 $^{(3)}$ & 0.12$ \pm $0.07 $^{(3)}$   \\
$V_{max}$ [km/s] & 215.8$ \pm $6.7 & -  \\
$r^{\prime}$-$g^{\prime}$ & -1.55$ \pm $0.04 & -1.50$ \pm $0.04  \\
$B$ & -21.69$ \pm $0.13 & -17.43$ \pm $0.15  \\
$L_{B}$ [$L_{\odot}$] & (6.9$ \pm $1.3)$\times$10$^{10}$ & (6.1$ \pm $0.9)$\times$10$^{8}$ \\ %
$M_{HI}$ [$M_{\odot}$] & (8.3$ \pm $0.4)$\times$10$^{9}$ & -   \\
$M_{dyn}$ [$M_{\odot}$] & (3.3$ \pm $0.4)$\times$10$^{11}$ & -   \\
$M_{HI}$/L$_{B}$ [$M_{\odot}$/$L_{\odot}$] & 0.12$ \pm $0.01 & - \\
\hline
\end{tabular}
\end{table}

\section{Gemini observations}

\par Optical observations were conducted at the Gemini-North telescope using the imaging and
multi-slit spectroscopic capabilities of the Gemini Multi-Object Spectrograph (GMOS) in
queue mode.

\par Before the observations, images of NGC 6907 publicly available were used to pre-select the
best H~{\sc ii} regions candidates on it as well to determine the optimum position angle for
the observation ($243^\circ$ or $63^\circ$ - from north to east). This angle minimize the
number of selected objects that would have overlapping spectra on the GMOS detector.

\par As frequently happens, the MOS observations were broken in two stages: one for
imaging and the other for spectroscopy. The images obtained in the first stage were employed
to design the metallic mask with slits to perform the spectroscopy of the field and then
used for photometric and morphological studies.

\par As required in the queue mode a set of minimum conditions under which the observation
could be executed were solicited. A minimum image quality of $70\%$ was requested and for
this reason observations at 2$\times$2 binned mode were chosen, shortening the readout time.
As a compromise between spectral resolution and amount of light detected all the slits were
fixed with 1 arcsec widths. A standard slit height of 5 arcsec was chosen to improve the
spatial sampling of the sky, since the typical source size of the candidates is almost 2
arcsec.

\par All the observations were executed in better conditions than required, ensuring
full photometric conditions in all images. The seeing was 0.7 arcsec and the sky brightness
was as good as 21.7 mag arcsec$^{-2}$ in the g$^{\prime}$ filter and 20.1 mag arcsec$^{-2}$
for the r$^{\prime}$ filter.

\par For all the observations fast tip-tilt guiding was performed using wavefront sensors for
image motion compensation. Only one on-instrument wavefront sensor (OIWFS) guide star
(GSC0691101076) was needed for both imaging and spectroscopy observations.

\par Optical imaging for the NGC 6907 field was executed on August 1st 2005 using the
filters g$^{\prime}$ (g\_G0301) and r$^{\prime}$ (r\_G0303). To guarantee a continuous
coverage of the NGC 6907 field we applied different dithering patterns to integrate a single
frame in each filter.

\par
\begin{figure}
\includegraphics[width=86mm]{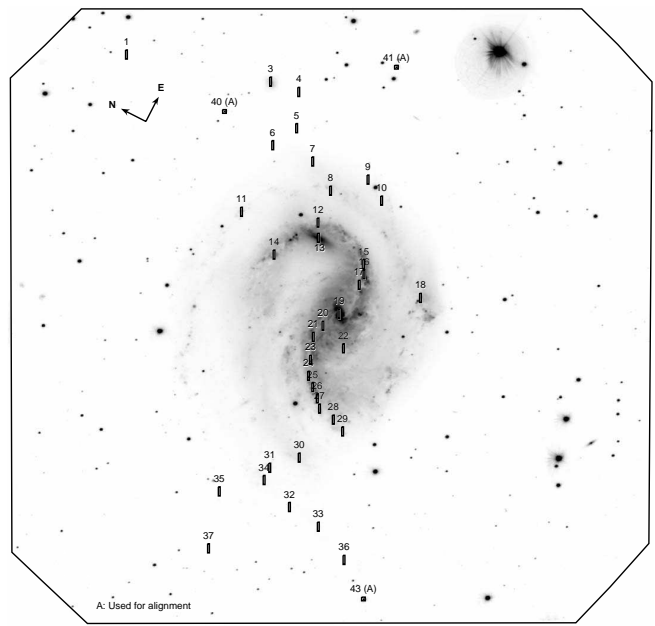} \caption{GMOS image of
NGC 6907 in the g$^{\prime}$ filter overlapped by the representation of the slits. The slit
13 was used to sample NGC 6908 and the slit 19 to sample the optical center of NGC 6907.
Special squared slits signed with (A) were used to guarantee the alignment of the slits over
the scientific targets.} \label{fig2}
\end{figure}

 The image files prepared in the first stage of the observations were used to select 33
H~{\sc ii} regions candidates over the NGC 6907 field, according to their brightness and
geometry (Figure \ref{fig2}). Standard procedures using the GMOS Mask Making Software were
employed to design a mask with slits over each one of these candidates. A special slit was
reserved for NGC 6908 (Slit 13).

\par The spectroscopic observations were carried out in September 2nd 2005 in a single
exposure of 960 seconds using the B600$\_$G5303 grating. This disperser and the slit width
of 1 arcsec guarantees a spectral resolution of $\sim$5 {\AA}. No spectral dithering was
used to cover the gaps on the final spectra, so special care was taken in choosing the
central wavelength of 5000 {\AA} by considering the systemic velocity of NGC 6907 and its
expected rotation curve to avoid signal losses. The final spectra covered wavelengths
between 3500 {\AA} - 6300 {\AA}. A CuAr arc lamp and a flat-field frame were observed in the
sequence of the observation, while the telescope was still following the source, to
guarantee the quality of the calibration frames, by avoiding any effect caused by torques
over the instrument.

\subsection{Data reduction procedures for Gemini Observations}

\par There are great differences for imaging and spectroscopy GMOS reduction, but for both
cases the Gemini Data Reduction Software was used.

\par The reduction for images in filter g$^{\prime}$ and r$^{\prime}$ followed the same
procedures. Individual dithered images of the field were corrected by overscan level,
differences in gain for the detectors, bias and flatfield. After these steps images from
different detectors were mosaiced and all exposures were co-added in a clean image.

\par Spectroscopic data reduction requires some extra cares. The first step was to prepare
combined and normalized flatfield images, correcting this frames by overscan and bias. Since
there was offsets between flatfield and the acquisition frames the normalized flatfield was
used for finding the slit edges using the gradient method. For this observation, an offset
of 6 pixels in the spatial direction was applied.

\par Considering that a single frame was acquired in the spectroscopic observations, the
cosmic rays could not be removed by combining frames. In this case the task GSCRREJ was
used. Residual effects on the spectral lines were corrected reprocessing only the slits
affected by such correction, making the parameters for the cosmic ray removing more
flexible.

\par An interactive process for the identification of spectral lines was set in each
one of the spectrum analyzed. Multiple solutions were obtained for the same spectrum using
the spatial distribution of the spectral lines, improving the wavelength calibration.

\par For each slit a local sky spectrum was extracted to be used to subtract the sky
lines and continuum from the target spectrum. Standard sky lines were used to verify if
there was raw systemic error in the observation.

\par After the visual inspection, the final spectra were generated by considering the
on source integrated emission in the spatial direction.

\section{GMRT observations}

\par The interferometric mode of the Giant Meterwave Radio Telescope (GMRT) was used for
21 cm imaging and spectroscopy. A more detailed description about this instrument is given
by Swarup et al. (1991) and Ananthakrishnan \& Pramesh Rao (2002).

\par The observations were carried out with a total baseband bandwidth of 4.0 MHz
divided into 128 channels centered at 1405.40 MHz to cover the H~{\sc i} heliocentric
velocity range of 2755 - 3610 km s$^{-1}$ (optical definition). A summary of the main
observational parameters are presented in Table \ref{tbl-2}.

\par The spectral resolution of the observation was 6.7 km s$^{-1}$ per channel.
Topocentric to heliocentric conversion of frequency was done with the help of DOPSET task to
take care of the earth's motion at the date and time of observation.

\par The radio sources 3C286 and 3C48 were observed at the beginning and the end of the
observation run as flux calibrator for $\sim$ 11 minutes each. VLA calibrator source
1923$-$210, which is an unresolved point source for GMRT (flux density of 2.0 Jy at 20 cm)
was used as the phase calibrator. It was observed alternately for $\sim$ 5 minutes every
$\sim$ 20 minutes observing NGC 6907.

\begin{table}
\caption{Observational parameters for the GMRT observation.}
 \label{tbl-2}
 \centering
 \begin{tabular}{lr}
 \hline
{Parameter} & {Value}\\
\hline
Date of Observation & 2006 July 13 \\ %
Pointing center (RA J2000) & 20$^{h}$25$^{m}$06$^{s}$.652 \\
Pointing center (DEC J2000) & -24$^{\circ}$48$^{\prime}$33$^{\prime\prime}$.492 \\
Observing duration[hrs] & 4.7 \\
Total on source time [hrs] & 2.6\\
Central frequency [MHz] & 1405.40\\
Bandwidth per IF [MHz] & 4.0 \\
Number of spectral channels & 128 \\
Polarizations & 2 \\
Frequency resolution [kHz] & 31.3 \\
Velocity resolution [km.s$^{-1}$] & 6.7 \\
Flux calibrators & 3C48 and 3C286 \\
Phase calibrator & 1923-210 \\
Bandpass calibrator  & 3C48, 3C286 and 1923-210 \\
\hline
\end{tabular}
\end{table}

\subsection{Data reduction procedures for GMRT Observations}

\par The radio data were reduced following standard calibration and imaging methods
employing the classical AIPS. Flux densities at the observing frequency were estimated using
the standard primary calibrators 3C286 and 3C48, also used to evaluate the flux densities
and phases of the secondary calibrator. For 3C286 the estimated flux density was 14.81 Jy
and for 3C48 it was 16.03 Jy. Taking into account the results of Omar et al.(2002) for 3C48,
the accuracy in the flux calibration must be better than $3\%$. Since there was not any
spectral features for 3C48, 3C286 and 1923$-$210 at the observing frequency, all of them
were used to bandpass calibration.

\par Standard spectral line data reduction procedure was followed to remove bad data, to
calibrate the gains and bandpass to produce the final calibrated visibilities for the
target field.

\par A preliminary datacube was generated at spatial frequencies below 5 k$\lambda$ to
verify which channels would contain H~{\sc i} emission. Visual inspection, channel by
channel, enabled to register the detection of extended H~{\sc i} emission within the
channels 41-105.

\par The polyhedron imaging technique was used to generate continuum images by averaging
42 line-free channels (ranging from 7 to 36 and 112 to 125), with data self-calibrated in both
phase and amplitude. Spatial frequencies below $5 k\lambda$ were used and a Gaussian taper
function was applied to weight down long baselines, reducing the sidelobes. In this case the
overall noise level was 2 mJy. The continuum flux density was subtracted from each
individual channels and the final self-calibrated solutions were applied to the line-data to
make an image cube with a resolution of 52$^{\prime\prime} \times$ 39$^{\prime\prime}$. To
enhance the sensitivity to extended features the data points were "natural-weighted" and the
resulting spectral cubes were cleaned for signals greater than 2.5 times the rms noise in
the channel images.

\par To improve the spatial resolution for the analysis, several iterations of self-
calibration were executed to make image cubes at different UV-ranges adapting the
previous procedure.

\par The relationship between frequency and spectral-line velocity was set by means of the
standard task ALTSWTCH. Moment maps were made from the resulting data cubes by using the
AIPS task MOMNT (see the contours in Figures \ref{fig6} and \ref{fig9} for the 0th and 1st
moment map respectively).

\section{Results}

\subsection{Optical Photometry}

\par The photometric calibration for the images in the filters g$^{\prime}$ and r$^{\prime}$
was made using the photometric standard stars in the MARK-A field (Landolt 1992), supplied
by the Gemini baseline (GN-CAL20050611) and the usual procedures of the DAOPHOT IRAF tasks.
Solutions including the atmospheric extinction were considered to determine the zero point
magnitudes for the g$^{\prime}$ (m$_0$ = 31.72) and r $^{\prime}$ (m$_0$ = 31.60). The
results are in excellent agreement with the median atmospheric extinction coefficients for
Mauna Kea, with differences lower than $0.4\%$

\par The limit of detection of the galaxies was taken as the $3\sigma$ limit of dispersion of
the local sky level. For NGC 6907 these limits are 24.7 $\pm$ 0.2 mag arcsec$^{-2}$
(g$^{\prime}$) and 23.6 $\pm$ 0.2 mag arcsec$^{-2}$ (r$^{\prime}$). For NGC 6908 they are
22.6 $\pm$ 0.3 mag arcsec$^{-2}$ and 21.4  $\pm$ 0.3 mag arcsec$^{-2}$ respectively to the
same filters.

\par Elliptical fittings in these isophotal limits, masked from the contribution of stellar
fields, the interarm regions and the disturbed parts of the galaxy provided the projection
parameters of NGC 6907 and 6908 (Table \ref{tbl-3}).

\begin{table}
%\begin{minipage}{126mm}
 \caption{Parameters derived from the elliptical fitting to optical and radio isophotal data
at 3$\sigma$ of the sky or noise level; $a$ and $b$ are the major and minor axis of the
ellipse, in arcsec, $\alpha_0$ and $\delta_0$ are the equatorial coordinates for the center
(J2000), $i$ is the inclination, $PA$ is the position angle, both in degrees.}
 \label{tbl-3}
 \begin{tabular}{llll}
 \hline
{Para-} & {NGC 6907} & {NGC 6907} & {NGC 6908} \\
{meter} & {Optical $\langle r^{\prime}$+$g^{\prime} \rangle$} & {Radio (5k$\lambda$)} & {Optical $\langle r^{\prime}$+$g^{\prime} \rangle$} \\
 \hline
$a$ & 88.2$ \pm $1.8 & 212$ \pm $19 & 10.5$ \pm $0.8  \\ %
$b$ & 77.4$ \pm $1.2 & 148$ \pm $10 & 3.2$ \pm $0.2    \\
$\alpha_0$ & 20$^{h}$25$^{m}$06$^{s}$.29 & 20$^{h}$25$^{m}$08$^{s}$.70 & 20$^{h}$25$^{m}$08$^{s}$.96  \\
$\delta_0$ & -24$^{\circ}$48$^{\prime}$16$^{\prime\prime}$.3 & -24$^{\circ}$48$^{\prime}$43$^{\prime\prime}$.6 & -24$^{\circ}$48$^{\prime}$03$^{\prime\prime}$.4 \\
$i$ & 29.1$ \pm $4.2 & 46.4$ \pm $3.1 & 74.1$ \pm $4.5 \\
$PA$ & 74.1$ \pm $6.6 & 52.3$ \pm $11.5 & 3.1$ \pm $2.1 \\
 \hline
\end{tabular}
%\end{minipage}
\end{table}

\par The integration of all flux inside these isophotes in each filter gives the g$^{\prime}$
and r$^{\prime}$ magnitudes. These values were converted to the standard B-magnitude using
the expressions by Fukugita et al. (1996) and Hook et al. (2004), corrected by the foreground
Galactic extinction (Am{\^o}res \& L{\'e}pine 2005) and masked from the foreground
contributions. The results were a total B-magnitude of $11.55 \pm 0.03$ and a mean effective
surface brightness of 22.36 $\pm$ 0.05 mag arcsec$^{-2}$ for NGC 6907 and a total B-magnitude
of 15.81 $\pm$ 0.15 and a mean effective surface brightness of 20.81 $\pm$ 0.08 mag
arcsec$^{-2}$ for NGC 6908.

\par
\begin{figure}
\includegraphics[width=84mm]{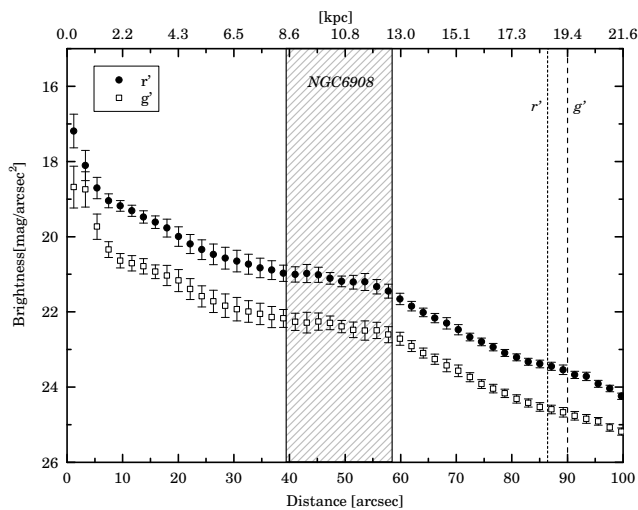} \caption{g$^{\prime}$ and r$^{\prime}$ radial surface
brightness profiles. Special limits mentioned in the text are indicated by vertical or
horizontal lines over the graph.}
 \label{fig3}
\end{figure}

 The radial surface brightness profiles were obtained performing the integration of
the flux on rings with the same size of the seeing (0.7 arcsec) and following the parameters
of projection observed for the galaxy in each filter. For NGC 6907 (Figure \ref{fig3}) the
central variation of the brightness is compatible with a small spiral structure (usually
identified as a bar) within (7.4 $\pm$ 2.1) arcsec. The exponential behavior is only
affected at radius near NGC 6908 emission, returning back to the exponential behavior for
larger distances. For NGC 6908 the surface brightness profiles are compatible with Madore et
al. 2007, but the extended central emission of NGC 6908 suggests the presence of a composed
nucleus.

\subsection{Optical Spectroscopy}

\par For the spectrophotometric calibration the sensitivity function was achieved using the
flux standard star G191B2B supplied by the Gemini baseline, according to the specifications
for the NGC 6907 spectral range coverage.

\par The IRAF package RVSAO was used to identify the spectral lines observed in the H~{\sc ii}
regions of NGC 6907 by cross-correlating a template spectrum of this kind of object to the
observed spectra. A set of shifts and errors were determined for each spectral line and the
radial velocities of all lines in each spectrum were combined by discarding the weakest
lines and weighting the velocities with their errors and equivalent widths. The results of
these process are summarized in Table \ref{tbl-4}.

\begin{table}
 \caption{Observed radial velocities in direction of the selected H~{\sc ii} regions in
the optical (v$_{opt}$) and in 21cm (v$_{HI}$). The ID is the slit number, $RA$ and $Dec$
are the equatorial coordinates (J2000), $r$ is the de-projected distance from the center,
considering the projection parameters of the radio for which the uncertainties are lower
than 0.2 kpc ($H_0 =73 \pm 5)$ km s$^{-1}$/Mpc.}
 \label{tbl-4}
 \begin{tabular}{ccccccc}
 \hline
{ID} & {RA} & {Dec} & {r} & {v$_{HI}$} & {v$_{opt}$} \\
&[$^{h}$]&[$^{\circ}$]& [kpc] & [km/s] & [km/s] \\
 \hline
19 & 20.418508 & -24.809166 & 0.8 & 3189$ \pm $92 & 3209$ \pm $10  \\ %
20 & 20.418309 & -24.807866 & 2.9 & 3234$ \pm $87 & 3182$ \pm $18  \\
22 & 20.418197 & -24.812122 & 4.5 & 3299$ \pm $32 & 3366$ \pm $28 \\
21 & 20.418173 & -24.807157 & 4.9 & 3224$ \pm $86 & 3181$ \pm $22  \\
17 & 20.418904 & -24.809892 & 5.6 & 3055$ \pm $56 & 3048$ \pm $9  \\
16 & 20.418995 & -24.810027 & 6.8 & 3098$ \pm $63 & 3136$ \pm $6  \\
23 & 20.417903 & -24.808552 & 7.7 & 3264$ \pm $62 & 3224$ \pm $11  \\
15 & 20.419107 & -24.809228 & 7.7 & 3055$ \pm $54 & 3123$ \pm $9  \\
13 & 20.419157 & -24.801142 & 8.8 & 3013$ \pm $37 & 3092$ \pm $52 \\
24 & 20.417721 & -24.809431 & 9.7 & 3295$ \pm $37 & 3350$ \pm $28  \\
25 & 20.417622 & -24.810875 & 10.7 & 3303$ \pm $31 & 3316$ \pm $11 \\
12 & 20.419317 & -24.799984 & 10.7 & 3000$ \pm $32 & 2948$ \pm $9 \\
26 & 20.417594 & -24.811851 & 10.9 & 3306$ \pm $30 & 3314$ \pm $20 \\
14 & 20.418777 & -24.796278 & 11.5 & 3044$ \pm $40 & 3062$ \pm $16 \\
27 & 20.417468 & -24.813068 & 12.4 & 3312$ \pm $25 & 3323$ \pm $11 \\
28 & 20.417435 & -24.815659 & 13.1 & 3315$ \pm $24 & 3363$ \pm $66 \\
29 & 20.417363 & -24.817730 & 14.3 & 3316$ \pm $23 & 3365$ \pm $25 \\
08 & 20.419713 & -24.799416 & 14.6 & 2996$ \pm $34 & 2982$ \pm $10 \\
18 & 20.419087 & -24.819066 & 14.9 & 3178$ \pm $37 & 3140$ \pm $5  \\
10 & 20.419847 & -24.807111 & 16.3 & 3045$ \pm $55 & 3081$ \pm $25  \\
09 & 20.419967 & -24.803985 & 17.2 & 3034$ \pm $55 & 2969$ \pm $26  \\
11 & 20.419039 & -24.788920 & 18.5 & 3041$ \pm $29 & 3024$ \pm $9  \\
31 & 20.416655 & -24.810240 & 22.9 & 3250$ \pm $21 & 3277$ \pm $19 \\
34 & 20.416497 & -24.810383 & 24.8 & 3240$ \pm $14 & 3263$ \pm $38 \\
35 & 20.416154 & -24.805107 & 31.0 & 3231$ \pm $15 & 3201$ \pm $20 \\
 \hline
\end{tabular}
\end{table}

\begin{figure}
\includegraphics[width=86mm]{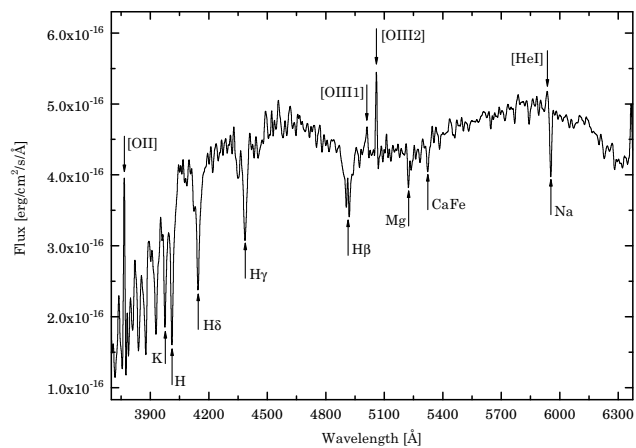} \caption{NGC 6908 integrated spectrum.
smoothed by removing Fourier components of higher frequencies. Absorption and emission lines
are identified by the arrows.}
 \label{fig4}
\end{figure}

\par
\begin{figure}
\includegraphics[width=86mm]{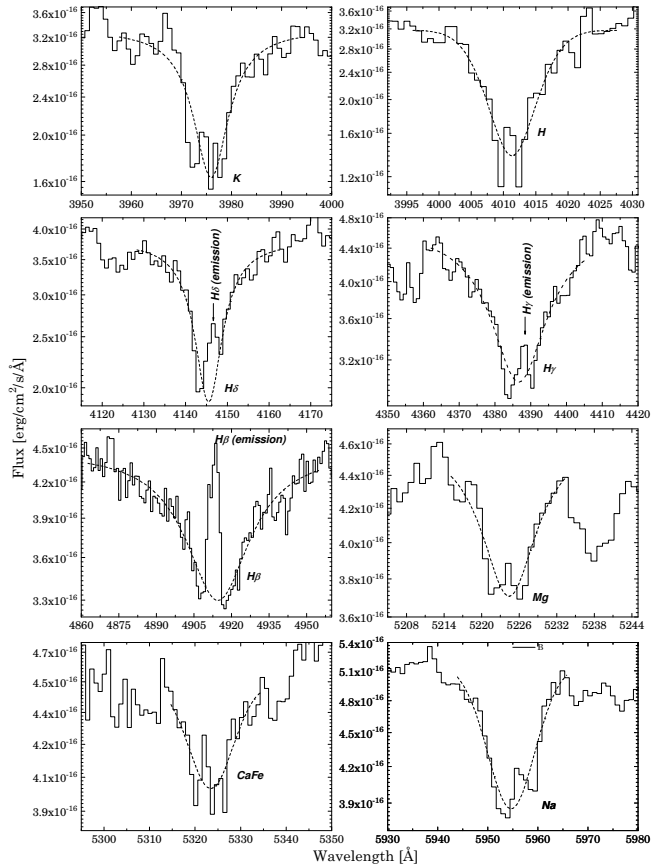} \caption{Detail of each
absorption line (continuous line) and the fitted profile (dashed line). A Gaussian or
Lorentzian profile was chosen depending on which minimizes the $\chi^{^2}$ per degree of
freedom.}
 \label{fig5}
\end{figure}

 Special care was taken with the NGC 6908 spectrum (Figure \ref{fig4}), since its optical depth is unknown
and it can have contributions coming from NGC 6907 in its own spectrum. Taking into account
that NGC 6908 is a lenticular S0 galaxy, then absorption lines must provide the most reliable
information about its contents. Visual inspection was done to recognize spectral absorption
lines and individual fitting profiles were performed considering Gaussian and Lorentzian
models to obtain the central wavelength of the lines (see Table \ref{tbl-5}). Emission lines
were detected inside the absorption lines, specially in the hydrogen lines (Figure
\ref{fig5}), confirming the spectra superposition.

\begin{table}
 \caption{Central wavelengths and velocities for the absorption spectral lines
observed in direction of NGC 6908. $\lambda_{0}$ is the rest wavelength. $\lambda_{abs}$ is
the peak centroid, and $v_{abs}$ is the velocity associated to displacement of the observed
spectral line.}
 \label{tbl-5}
 \centering
 \begin{tabular}{lccc}
 \hline
{Absorption } & {$\lambda_{0}$} & {$\lambda_{abs}$} & {$v_{abs}$} \\
Line& [{\AA}] & {[\AA]} &  [km/s] \\
 \hline
K & 3933.70 & 3975.91$ \pm $0.36 & 3217$ \pm $13 \\ %
H & 3968.50 & 4011.31$ \pm $0.28 & 3234$ \pm $16 \\
H$\delta$ & 4101.70 & 4145.49$ \pm $0.24 & 3201$ \pm $28 \\
H$\gamma$ & 4340.50 & 4386.68$ \pm $0.31 & 3189$ \pm $28 \\
H$\beta$ & 4861.33 & 4914.56$ \pm $0.54 & 3283$ \pm $21 \\
Mg & 5175.36 & 5224.40$ \pm $0.28 & 2841$ \pm $17 \\
CaFe & 5268.98 & 5323.46$ \pm $0.49 & 3099$ \pm $21 \\
Na & 5892.50 & 5954.66$ \pm $0.25 & 3162$ \pm $33 \\
\hline
\end{tabular}
\end{table}

\subsection{Continuum}

\par A more detailed view of the continuum emission of NGC 6907 shows a north-east elongated
structure, with an asymmetric emission which does not provide good elliptical fitting
(Figure \ref{fig1}). The interaction between NGC 6907 and 6908 must be the main reason for
such asymmetry. Almost all the continuum emission is confined to an isophote of 1.9 mJy/beam
with approximately the same size than the optical emission centered at
$\alpha_{c}$=20$^{h}$25$^{m}$07$^{s}$.28 and
$\delta_{c}$=-24$^{\circ}$48$^{\prime}$33$^{\prime\prime}$.9 .  The total continuum flux
measured for NGC 6907 is 112 mJy, in absolute agreement with Condon et al. (1998) and Vollmer
et al. (2005), if the same isophotal limit is considered. The radial profile of this
emission exhibits the expected behavior for an exponential disk inside a de-projected radius
of 72 arcsec.

\subsection{H~{\sc i} Emission}

\par Subtracting the continuum image from the image cubes composed for UV-ranges up to 40
k$\lambda$ it was possible to recover the spatial H~{\sc i} distribution for each channel,
resolving structures larger than 10$\times$12 arcsec. That is enough to sample the
integrated contribution of the components inside the galaxy NGC 6907.

\par
\begin{figure}
\includegraphics[width=86mm]{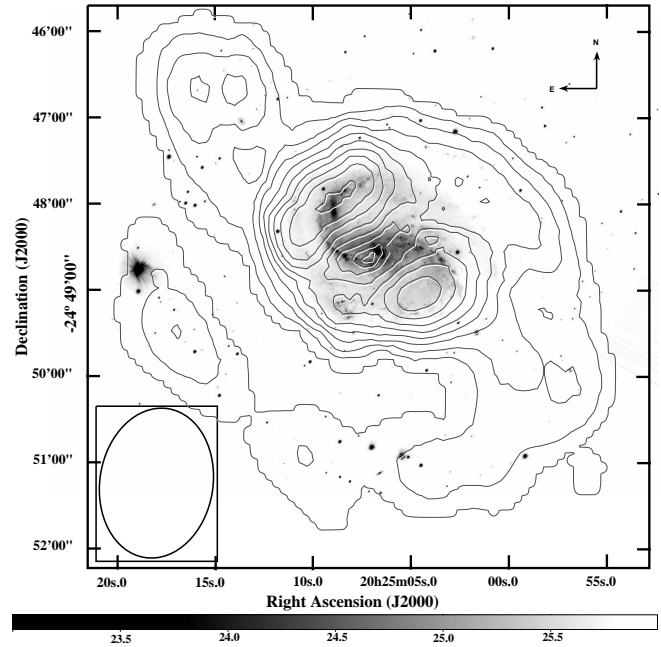}
 \caption{{H~{\sc i}} column density map over the
g$^{\prime}$ image of NGC 6907 and 6908. The contour levels are at 0.3, 8.8, 17.3, 25.9,
34.4, 42.9, 51.4, 60.0, 68 .5 and 77.0 in units of $10^{19} cm^{-2}$.}
 \label{fig6}
\end{figure}

 The bidimensional integration of the H~{\sc i} emission detected above $2\sigma$ of
the noise level in all channels composes the 0th moment. In Figure \ref{fig6} this map was
converted to the H~{\sc i} column density assuming the relation by Spitzer (1978). It
reveals a substantially more extended spiral structure than observed in optical images and
the emission has two extended peaks coincident with the spiral arms.

\par Elliptical fitting for the isophote associated with a column density of 2.8$\times
10^{19}cm^{-2}$ provides the extension of the radio emission as (3.5 $\pm$ 0.3) arcmin or
almost 2.4 times larger than the optical emission (Table \ref{tbl-3}). Neither the optical
center nor the radio center are coincident.

\par
\begin{figure}
\includegraphics[width=86mm]{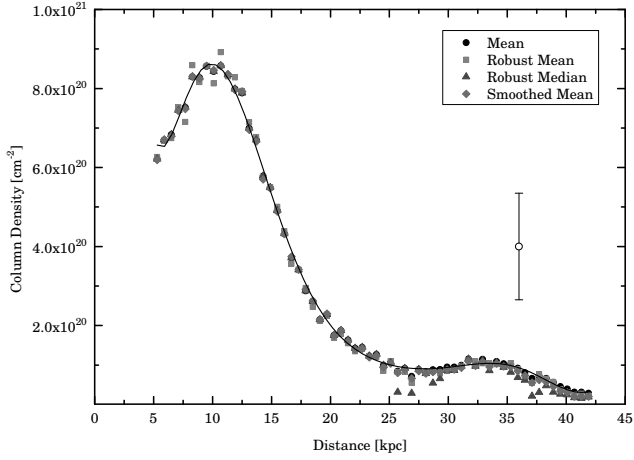} \caption{Radial H~{\sc i} column density
 profile. The open circle with error bars informs the mean error in data.}
 \label{fig7}
\end{figure}

 The integration of the column density on elliptical rings along NGC 6907 provides the
radial column density profile presented in Figure \ref{fig7}. As expected, there is
depletion of H~{\sc i} on the central regions of NGC 6907, since the molecular hydrogen is
formed in a denser environments. After the peak emission at 10 kpc an exponential fall
occurs up to 28 kpc, where the column density rises up to 34 kpc and falls back again. Such
distribution is quite common as can be seen in Broeils \& van Woerden (1994), and the
minimum inside the galactic disk can be related to the gas redistribution in the corotation
radius.

\par
\begin{figure}
\includegraphics[width=86mm]{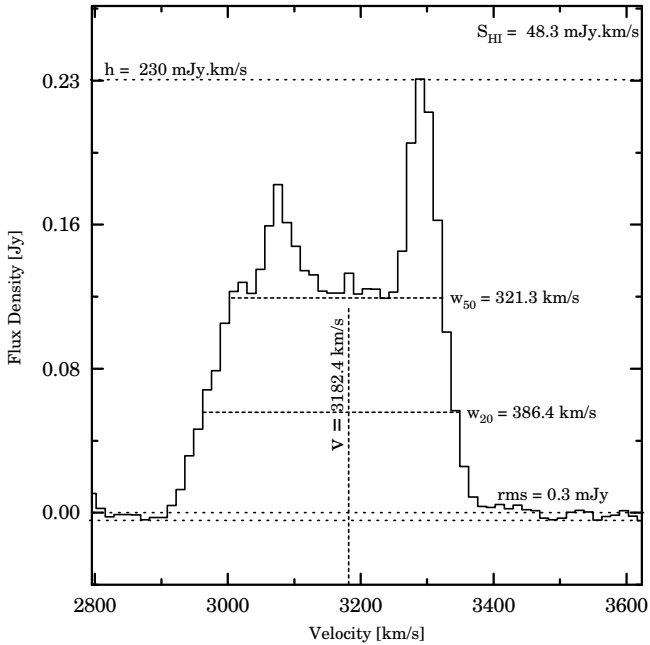} \caption{Global H~{\sc i} emission profile of NGC 6907 and 6908.
The bottom dotted lines limit the noise level while the top dotted line indicates the
maximum emission $h$. Inside the distribution two horizontal dashed lines are used to mark
the H~{\sc i} line-widths estimated at 20\% (w$_{20}$) and at 50\% (w$_{50}$) of the peak
intensities; $v$ is the central weighted mean velocity distribution.}
 \label{fig8}
\end{figure}

 On the other hand, the integration of all the emission for each channel of the datacube
after continuum subtraction provides the global H~{\sc i} emission profile for the galaxies
NGC 6907 and 6908 as it can be seen in Figure \ref{fig8}.

\par The double peaked emission is expected for rotating disks. Nevertheless, the different
levels of the peaks indicates the presence of other components in the distribution of
velocities measured in the field. These effects can be attributed to the spatial velocity of
NGC 6908 projected on the line-of-sight and to the intrinsic asymmetries on the velocity
field of NGC 6907.

\par Richter \& Huchtmeier (1991), Chengalur et al (1993) and Doyle et al (2005) observed
these galaxies in lower resolution with different single dishes, what can possibly account
for the main differences in the registered line profiles. The mean differences are lower
than $3\%$, including the flux density (48.3 $\pm$ 1.8 mJy.km s$^{-1}$), suggesting that a
beam smearing is the main responsible for the differences found at the flux peak.

\par The weighted mean velocity associated with the H~{\sc i} profile ($3182.4 \pm 3.9$ km 
s$^{-1}$) is absolutely compatible with the velocity measured by the robust mean of the entire 
1st moment map (3183.2 $\pm$ 6.7 km s$^{-1}$). Taking into account the Hubble flow distance 
(supposing H = (73 $\pm$ 5) km/sec/Mpc), the galactocentric distance of this galaxy is estimated 
as (44.5 $\pm$ 3.1) Mpc. Using this distance a  H~{\sc i} mass of (8.3 $\pm$ 0.4)$\times 10^9$ 
$M_{\odot}$ was calculated.

\par
\begin{figure}
\includegraphics[width=86mm]{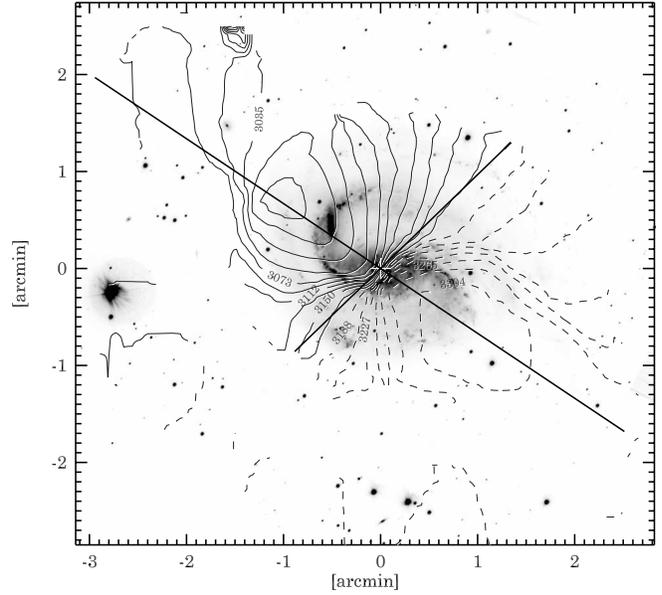} \caption{NGC 6907 velocity field
superimposed over its image in g$^{\prime}$. Dashed lines represent the part of the velocity
field going away from the observer while the opposite is represented by the continuous
lines. The thicker crossing lines represent the kinematical axis. The velocity field was
generated using an UV range of $5k\lambda$ and a resolution of $52^{\prime\prime} \times
39^{\prime\prime}$, to balance the sensitivity to the diffuse emission and the spatial
resolution. }
 \label{fig9}
\end{figure}

 Considering the distribution of velocities associated with the bidimensional emission
in each frequency of the 1st moment map, then the observed velocity field of the galaxy can
be extracted (Figure \ref{fig9}).

\par One can note that the velocity field is strongly oriented by the spiral arm in the
internal regions. In fact, the high density of the H~{\sc i} emission in the spiral arms
(from 10 to 100 times higher than the interarm regions) could generate a bias in the
observed velocity field, if the velocity from the disk does not correspond to that measured
in the spiral arms.

\par The kinematic major axis is not perpendicular to the kinematic minor axis (Figure
\ref{fig9}). That can be easily verified by comparing the linear fit over the coordinates of
the points with the systemic velocity (minor axis) and the coordinates of the nearest points
to the minor axis of all isovelocity levels, which samples the major axis in an axisymetric
rotation disk. The angle between these axis is (78.2 $\pm$ 3.1)$^{\circ}$ and the point
where they cross each other can be used to define the kinematic center
($\alpha_{k}$=20$^{h}$25$^{m}$06$^{s}$.40 and
$\delta_{k}$=-24$^{\circ}$48$^{\prime}$38$^{\prime\prime}$.3).

\par Since the the global H~{\sc i} profile suggest that a rotating disk must be the main
component of the movement of the system, then the rotation curve of the galaxy NGC 6907 can
be extracted supposing that the elliptical fitting made on the H~{\sc i} gas distribution
represent the projection of the NGC 6907 disk on the sky and its gas follows circular orbits.
In this situation the observed line-of-sight velocity ignoring vertical and radial movements
would be:

\bigskip
\begin{equation}
v(\alpha,\delta)=V_{0} +v_{rot}(r,\theta)\sin(i)\cos(\theta)
\label{eq3}
\end{equation}
\bigskip

taking into account that:

\bigskip
\begin{equation}
\cos(\theta)=\frac{-(\alpha-\alpha_{0})\sin(\phi)\cos(\delta)+(\delta-\delta_{0})\cos(\phi)}{r}
\label{eq4}
\end{equation}
\bigskip

\bigskip
\begin{equation}
\sin(\theta)=\frac{-(\alpha-\alpha_{0})\cos(\phi)\cos(\delta)-(\delta-\delta_{0})\sin(\phi)}{r\cos(i)}
\label{eq5}
\end{equation}
\bigskip

where $(\alpha_{0},\delta_{0})$ is the kinematic center, $V_{0}$ is the systemic velocity,
$\phi$ is the position angle of the receding semi-major axis, $i$ is the inclination between
the normal to the plane of the galaxy and the line-of-sight and $r$ and $\theta$ are the
polar coordinates inside the plane of the galactic disk.

\par Note that we introduce the term $\cos(\delta)$ multiplying all distances measured in
right ascension (what it is not usually made in the literature), otherwise it would
overestimate the galactocentric radius, depending on the declination of the source and the
position angle of each target relative to the center of the galaxy. For the NGC 6907 these
errors are predominantly of the order of 12\%.

\par
\begin{figure}
\includegraphics[width=86mm]{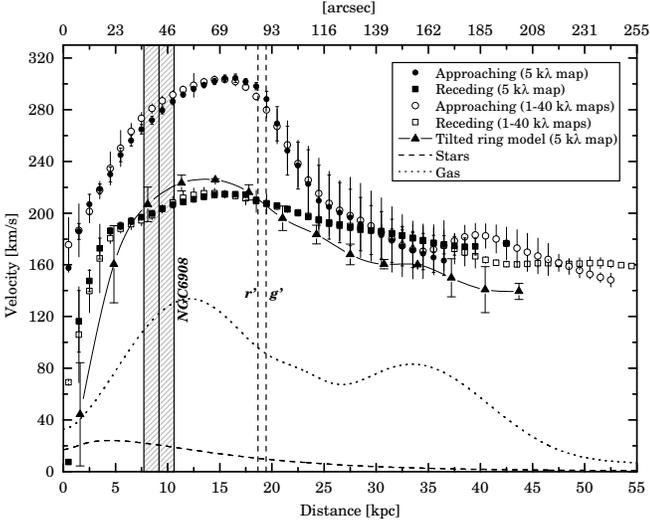} \caption{Rotation curves extracted
for the receding and approaching regions of the NGC 6907 velocity field. The hatched area
correspond to the positions occupied by NGC 6908 in the receding part of NGC 6907 and  two
vertical dashed lines represent elliptical fit limits in g$^{\prime}$ and r$^{\prime}$. For
the stellar rotation curve it was supposed a mass-to-light ratio of 1.5.}
 \label{fig10}
\end{figure}

 Using the combination of the velocity fields made with maps covering UV-ranges from
1k$\lambda$ to 40k$\lambda$, taking into account that the most reliable information about
the rotation curve is extracted near the kinematic major axis ($\sim \pm30^{\circ}$), then
the previous equations can be used to determine $v_{rot}$, $r$ and $\theta$ for the receding
and approaching regions of the galaxy (see Figure \ref{fig10}). The tilted ring model was
applied to the NGC 6907 velocity field, considering the same limits used in the previous
method. Variations of the inclinations and the position angles with the distance are lower
than the uncertainties for the position angle and inclination of the whole galaxy so its
results can be ignored. A raw estimate for the dynamical mass of NGC 6907, considering the
last point in the rotation curve is $(3.3 \pm 0.9)\times10^{11}$ $M_{\odot}$.

\par
\begin{figure}
\includegraphics[width=86mm]{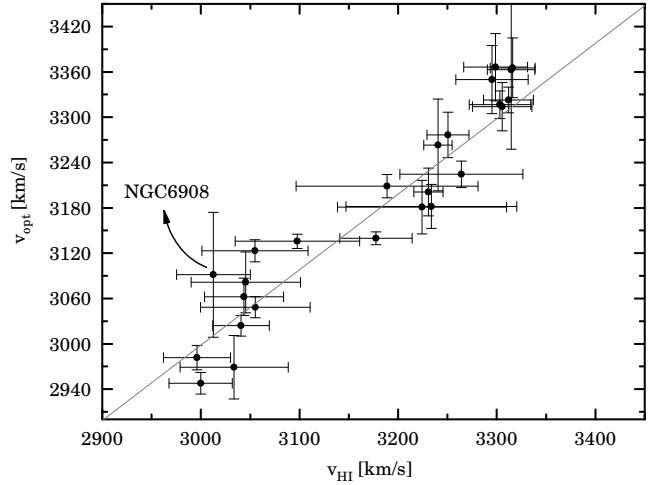} \caption{Comparison between
optical radial velocities observed with GMOS and radio radial velocities observed with the
GMRT. The straight continuous line represent the linear least squares regression for the
observed regions with azimuthal angles near the major axis ($\Delta\theta < 30^{\circ}$).}
 \label{fig11}
\end{figure}

 Comparing the radial velocities from optical and H~{\sc i} spectra observed in the same
directions (Figure \ref{fig11} and Table \ref{tbl-4} it is possible to verify the overall
correspondence between the optical and radio rotation curves $(v_{opt}=(1.00 \pm
0.08)v_{HI}+(1 \pm 240))$ and verify that the radio projection parameters provides the best
correspondences. The high dispersion of individual points is related to the higher
uncertainties to extract radial velocities azimuthaly far away from the major axis. In spite
of that, the mean dispersion is inferior to the limit of 50 km s$^{-1}$, refereed by Bosma
(1981) as the limit related to large-scale asymmetries of velocity fields. The exclusion of
the objects in larger azimuthal angles relative to the major axis $(\Delta\theta =
30^{\circ})$ provides almost the same fit. In particular, the point representing the slit which 
sampled NGC 6908 presents a higher velocity dispersion than the others, possibly due to the 
spatial motion of NGC 6908.

\section{Discussion}

\par A first view on the NGC 6907 velocity field (Figure \ref{fig9}) reveals that NGC 6908
is over the approaching part of it. The extraction of the rotation curve in this part of the
galaxy reveals contributions of radial components of the NGC 6908 movement, what does not
occurs in the receding part of the galaxy. This verification explains the different peak
levels in the global H~{\sc i} emission profile shown in Figure \ref{fig8} and in the
rotation curves presented in Figure \ref{fig10}. Taking into account that the H~{\sc i}
global profile preserves the behavior expected for a rotating disk and different emission
profiles agree with a exponential disk, then the effects of the interaction on the
velocities can be considered as local. Consequently the rotation curve extracted from the
receding part of the velocity field is the real (undisturbed) NGC 6907 rotation curve.

\par Comparing the raw observed rotation curves from the receding and approaching parts
of NGC 6907 in direction of NGC 6908, it is possible to see that the relative emission
velocity between these galaxies (130 $\pm$ 17 km s$^{-1}$, as measured by Madore et al.
2007) , coincides exactly with the expected velocity for the undisturbed rotation curve, and
not with the velocity measured in radio for the approaching velocity field (where NGC 6908
can be found). This apparent contradiction can be easily understood considering that the
emission line velocity measured for NGC 6908 is in fact from the rotating material in the
NGC 6907 disk. NGC 6908, as a lenticular galaxy, has less gas and new hot stars to heat it. So
the emission lines used by Madore to measure the NGC 6908 velocity only could come from the
material in the NGC 6907 disk.

\par The idea that the emission lines come from the NGC 6907 is corroborated by the
spectrum extracted in direction of NGC 6908 (Figure \ref{fig4}). It has a strong continuum
superimposed by few emission lines, for which the radial velocity is (3092 $\pm$ 52) km
s$^{-1}$, and by several absorption lines, associated to a radial velocity of (3147 $\pm$
21) km s$^{-1}$ (Table \ref{tbl-5}). Relative velocities spectroscopically measured in
relation to the  NGC 6907 center (3209 $\pm$ 10 km s$^{-1}$) are 117 $\pm$ 53 km s$^{-1}$ for
emission lines (compatible with the NGC 6907 rotation curve in that direction) and (62 $\pm$
23) km s$^{-1}$ for absorption lines, in perfect agreement with Madore et al. (2007).

\par Another confirmation about the origin of the emission lines is that the broaden
hydrogen absorption lines (typical of lenticular galaxies) contain vestiges of narrow
emission lines inside it (Figure \ref{fig5}). These emission lines have the same radial
velocity of the rotation curve of NGC 6907 at that radius. That is the reason why different
references and observations do not agree with the presence of certain hydrogen lines (they
can be hidden in lower resolutions observations) and why $H\beta$ is observed in absorption
while $H\alpha$ in emission. Since $H\alpha$ line is more intense than $H\beta$ the
equivalent absorption line formed in a different source is not deep enough to hide it.

\par
\begin{figure}
\includegraphics[width=86mm]{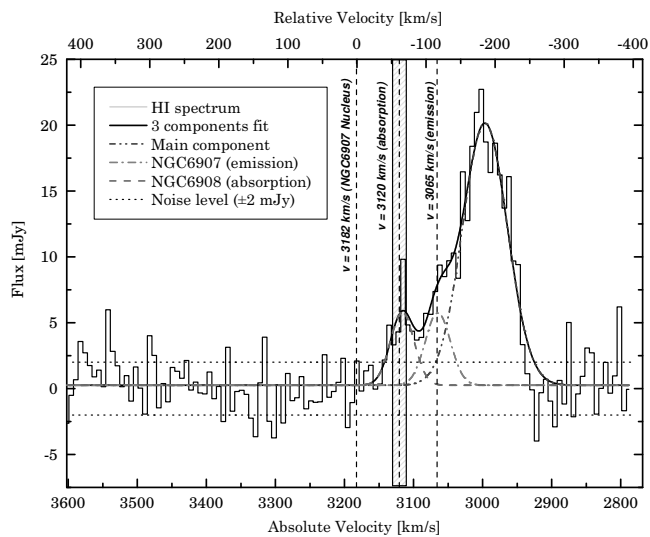} \caption{H~{\sc i} spectrum in direction
of NGC 6908 (gray stepped curve). The dashed vertical lines mark special velocities indicated
in the picture. The horizontal dotted lines limit the noise level. Three components were
fitted for this spectrum: the first associated with NGC 6908 by a dashed line, the second
related to the NGC 6907 disk emission lines (dot-dashed line) and the third related to the
gas left behind NGC 6908 due the interaction (dot-dot-dashed line). The sum of these
components is represented by the smoothed black curve overlaying the observed H~{\sc i}
profile (step gray line).}
 \label{fig12}
\end{figure}

 Verifying the distribution of H~{\sc i} emission as a function of the velocity in
the direction of NGC 6908, it is possible to distinguish three tied Gaussian components
(Figure \ref{fig12}). Considering the relative velocities measured with respect the radio
systemic velocity expected for the nucleus (since the spectrum extracted in direction of the
slit 19 does not sample the radio nucleus), the centroid of each component can be
interpreted as associated with different structures in the interaction between NGC 6907 and 6908. The first one, at 3120 km s$^{-1}$, is associated with NGC 6908, since it has the
same velocity of the absorption lines generated by its stellar content, the second one, at
3065 km s$^{-1}$, is produced by the excited material from the disk of NGC 6907 during the
interaction with NGC 6908 and the third one (main component) related to the higher velocity
gas left behind NGC 6908 due the interaction with NGC 6907.

\subsection{A scenario for the collision between the two galaxies}

\par
\begin{figure}
\includegraphics[width=86mm]{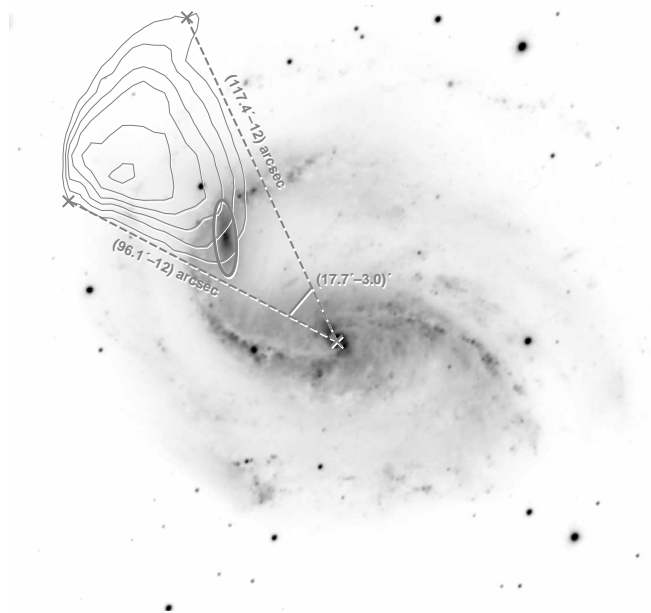} \caption{Contour of isovelocities
higher than 220 km s$^{-1}$ in the approaching region of the NGC 6907 velocity field
overlaying the g$^{\prime}$ image of this galaxy. The interval between contours is 5
km.s$^{-1}$.}
 \label{fig13}
\end{figure}

 Using the velocity contours above the approaching velocity of 220 km s$^{-1}$, it is
possible to trace the disturbed regions of the velocity field and see a converging
distribution of isophotes over NGC 6908, whose gradient is higher in direction of the
rotation of NGC 6907, following the spiral arms (Figure \ref{fig13}).

\par The striking triangular shape of the isophotes associated with high-velocity gas
could be explained by the following scenario. NGC 6908 crossed the disk of NGC 6907 in a
recent past, and very probably the tidal forces created a bridge of gas between the two
galaxies, similar to the Magellanic stream in our galaxy. The gas left behind NGC 6908 disk
by the collision explains the excess of gas approaching us. In our galaxy, at the solar
Galactic radius, the time required for a gas cloud launched in the direction perpendicular
to the plane to revert its motion and fall back onto it is about 35 Myr (see eg. L\'epine \&
Duvert, 1994). This can be taken as an order of magnitude for the present case. Let us now
consider the velocity component of NGC 6908 in the plane of the sky. This component is
directed approximately towards the center of NGC 6907, as we can infer from the fact that
presently NGC 6908 is situated closer to the center of NGC 6907 than the marks left in the gas
of the disk by the crossing event. The gas stream produced by the collision must also have a
component of velocity in the same (almost radial) direction, which explains why the
triangular isophotes point towards NGC 6908. Finally, let us analyze the component of
velocity of the stream in the plane of the sky, but in the direction of rotation of NGC 6907.
The stream can, in a simplified way, be considered as a mixture of gas with velocities
ranging from that of NGC 6908 (with almost zero component in the direction of rotation of
NGC 6907) to the gas which resulted to have about the same rotation velocity of the disk of
NGC 6907. This explains the elongated shape of the isophotes in the direction of rotation,
and at the same time, allows us to estimate in a new way the time elapsed since the
collision. The mean de-projected distance of the two extremes of the gas with perturbed
velocity in Figure \ref{fig13} is about $(104.8 \pm 17.9)$ arcsec, where the undisturbed
rotation curve has the velocity of $(200.0 \pm 6.7)$ km $s^{-1}$. Assuming $H_0 =(73 \pm 5)$
km s$^{-1}$/Mpc, this distance is $(22.6 \pm 4.1)$ kpc. Since the angle between the two extremes
is $(17.7 \pm 0.3)^{\circ}$, then the initial disturbed material would have traveled $(7.0
\pm 1.2)$ kpc azimuthally. From this we derive that the collision occurred about $(3.4 \pm
0.6)\times10^7$ years ago, which is consistent with order of magnitude previously estimated.

\section{Conclusions}

\par The combination of the GMRT radio observation and GMOS/Gemini optical observation
provides a more complete view about the interaction between the galaxies NGC 6907 and 6908.

\par Photometric images observed at the Gemini Telescope and the H~{\sc i} maps observed
at the GMRT reveals extended structures that can be understood as tidal debris in optical
images, but resemble the continuation of the spiral arms in radio. By fitting ellipses to
the last isophotes registered to each one of these observations for NGC 6907 we verified that
the 21-cm emission is 2.4 times more extended than the optical emission. The orientation of
the fitted ellipses  provide different  parameters of projection for optical and radio
observations, the differences being 15.3$^{\circ}$ for the inclination and 24.1$^{\circ}$
for the position angle. Geometric, photometric and kinematic centers are not coincident,
reinforcing the evidences about the effect of an interaction in NGC 6907.

\par The spectra observed using the GMOS spectrograph are coherent with those 
observed in 21-cm in the same direction of the slits. The parameters of projection in 
radio provides the best correspondences between the optical and radio velocities. The 
details of the emission and absorption lines observed in direction of NGC 6908 show 
that they have different origins. First, because the observed emission lines
are not typical of lenticular galaxies like [OII] (3727 {\AA}), [OIII] (4959 {\AA} and 5007
{\AA}) and the HeI (5876 {\AA}), specially for those with low luminosity like NGC 6908.
Second, because the radial velocity observed in absorption is different from the velocity
emission lines, which are coherent with the undisturbed rotation curve of NGC 6907 expected
for that direction. Finally, because inside the hydrogen absorption lines it is possible to
see traces of narrower hydrogen recombination lines with a Doppler effect suitable for the
NGC 6907 rotating disk, confirming their different origin.

\par The H~{\sc i} column density distribution presents a discontinuity in its radial
profile which can be can be related to the gas redistribution at the corotation  radius. The
details of the global H~{\sc i} emission profile reveals that the rotation is the main
component of the H~{\sc i} velocity distribution, and its asymmetry indicates the
contribution of the NGC 6908 motion is local and in velocities near to those found in the
NGC 6907 rotating disk. The total H~{\sc i} mass obtained was $(8.3 \pm 0.4)\times10^{9}$
$M_{\odot}$.

\par Due to the dominant contribution to the H~{\sc i} emission in the spiral arms
they can be the main contributors for the velocity field asymmetries, verified by the non
perpendicularity of the kinematic axis (difference of $12^{\circ}$).

\par The differences in the rotation curve extracted from the receding and
approaching parts of the velocity field can be interpreted as consequence of line-of-sight
components produced by the interaction with NGC 6908. Assuming that the interaction only
affected the side where NGC 6908 is observed in projection, and that the velocity field  of
the opposite side is the real rotation curve of NGC 6907, we were able to get a raw
estimative for the dynamical mass of $(3.3 \pm 0.4)\times10^{11}$ $M_{\odot}$.

\par Looking at the H~{\sc i} spectrum in direction of NGC 6908 it is possible to
distinguish three emission components: one coincident with the relative velocity for
NGC 6908, another coincident with the excited gas in NGC 6907 disk and the last one in
agreement with the higher relative velocities gas left behind NGC 6908 by dynamic friction.
Using the difference in the rotation curves extracted from the disturbed region of the
velocity field and that one away from this region, we found the limits for the disturbed
material on the NGC 6907. Supposing that this material has drifted inside the NGC 6907 disk,
according the rotation curve, we estimated the time when the interaction started in $(3.4
\pm 0.6)\times10^7$ years.

\par Another work is in preparation to discuss the metallicity distribution
in the NGC 6907 and its possible connection with the corotation radius.

\section{Acknowledgements}
\par We thank the staff of the GMRT that made these observations possible. GMRT is run by the
National Center for Radio Astrophysics of the Tata Institute of Fundamental Research.
Financial support for this work was provided by Conselho Nacional para o Desenvolvimento
Cient\'ifico e Tecnol\'ogico (CNPq) and Coordenac\~ao de Aperfei\c{c}oamento Pessoal de
N\'ivel Superior (CAPES). This research has been benefited by the NASA's Astrophysics Data
System (ADS) and Extra-galactic Database (NED) services. Their open software used in this
research is greatly acknowledged. The Gemini programme ID for the data used in this paper is
the GN-2005B-Q-39 (PI:S. Scarano Jr.). We thank C. Mendes de Oliveira and Jayaram N.
Chengalur for the helpful discussions.

%\newpage

\end{document}